\documentclass[fleqn,10pt]{wlscirep}
\usepackage[utf8]{inputenc}
\usepackage[T1]{fontenc}
\usepackage{lineno,hyperref}
\usepackage{comment}
\usepackage{float}
\usepackage{hyperref}
\usepackage{amsmath,amssymb,amsfonts}
\usepackage{array,multirow,makecell}
\usepackage{cleveref}
\usepackage{xcolor}
\usepackage{cite}
\usepackage{comment}
\usepackage[ruled,vlined,linesnumbered]{algorithm2e}
\usepackage{algorithmic}
\usepackage[justification=justified]{caption}
\usepackage{graphicx}
\usepackage{textcomp}
\usepackage{footmisc}
\usepackage{rotating}
\modulolinenumbers[5]
\title{Extracting Backbones in Weighted Modular Complex Networks}

\author[1,*]{Zakariya Ghalmane}
\author[2]{Chantal Cherifi}
\author[3]{Hocine Cherifi}
\author[1,4]{Mohammed El Hassouni}
\affil[1]{LRIT, URAC No 29, Rabat IT Center, University Mohammed V, Rabat, Morocco}
\affil[2]{DISP Lab, University of Lyon 2, Lyon, France}
\affil[3]{LIB EA 7534, University of Burgundy, Dijon, France}
\affil[4]{FLSH, Mohammed V University, Rabat 10500, Morocco}

\affil[*]{zakaria.ghalmane@gmail.com}


\begin{abstract}

Network science provides effective tools to model and analyze complex systems. However, the increasing size of real-world networks becomes a major hurdle in order to understand their structure and topological features. Therefore, mapping the original network into a smaller one while preserving its information is an important issue. Extracting the so-called backbone of a network is a very challenging problem that is generally handled either by coarse-graining or filter-based methods. Coarse-graining methods reduce the network size by grouping similar nodes, while filter-based methods prune the network by discarding nodes or edges based on a statistical property. In this paper, we propose and investigate two filter-based methods exploiting the overlapping community structure in order to extract the backbone in weighted networks. Indeed, highly connected nodes (hubs) and overlapping nodes are at the heart of the network. In the first method, called "overlapping nodes ego backbone", the backbone is formed simply from the set of overlapping nodes and their neighbors. In the second method, called "overlapping nodes and hubs backbone", the backbone is formed from the set of overlapping nodes and the hubs. For both methods, the links with the lowest weights are removed from the network as long as a backbone with a single connected component is preserved. Experiments have been performed on real-world weighted networks originating from various domains (social, co-appearance, collaboration, biological, and technological) and different sizes. Results show that both backbone extraction methods are quite similar. Furthermore, comparison with the most influential alternative filtering method demonstrates the greater ability of the proposed backbones extraction methods to uncover the most relevant parts of the network.

\end{abstract}
\begin{document}

\flushbottom
\maketitle
%
%


\section*{Introduction}
The complex networks paradigm offers great tools for modeling and understanding systems made of multiple interacting components \cite{cn1,cn2}. They are used by many researchers in numerous domains ranging from technology to biology \cite{ap1,ap2,ap3,ap4,ap5,ap6}. Extensive attention has been paid to reveal the common characteristics of complex networks, such as the scale-free behavior of the degree distribution \cite{scalefree}, the small-world property \cite{smallprop}  and the community structure \cite{c1,c2,c3,manish,moi3,moi4}. However, with the growing size of real-world systems, extracting relevant features and information becomes a much harder task to achieve. In such networks, valuable information is usually hidden by minor redundant intricacies. Therefore, it is of prime interested to devise effective methods in order to reduce the size of large-scale networks while preserving the information they carry. Extracting the backbone of the network can help to solve the conflict between the understanding of a network and its large size. Ideally, the backbone represents the core component of the network obtained by filtering all the redundant information. 
In recent years, an increasing interest has been witnessed in extracting backbones from large-scale networks. The existing methods can be divided into two classes: the coarse-graining methods and the filter-based methods. The coarse-graining methods \cite{back1,back2,boyack2005mapping} are based on the idea of grouping the nodes sharing some common characteristics. Then, they consider these groups as a single node, reducing the network size while preserving the properties of interest.  Within this framework, community detection is a common technique. In this case, communities are replaced by a single node. The common property shared by the regrouped nodes is that they have more links with nodes in their community than with nodes outside of their community. In  \cite{back3}, the authors define a coarse-graining method preserving random walks. As nodes sharing the same neighbors cannot be distinguished from the point of view of a random walk they are grouped together. Note, however, that there is no consensus on the properties that need to be preserved. The filter-based methods follow a bottom-up approach in order to extract the network backbone \cite{back4,zhang2014network,glattfelder2009backbone,zhang2013extracting,liebig2016fast,marotta2016backbone,cao2019motif,back5,carmi2007model,back6,serrano,li2014identifying,lindner2015structure,qian2015extracting,zhang2014extracting}. First, a node or an edge statistical property is defined. It is, then, used as a criterion to preserve or discard nodes or edges from the network. As a result, only nodes and links carrying relevant information according to the defined property are conserved while the rest is removed. Filter-based methods can be classified according to the type of information used to filter the nodes or edges (global, local, a combination of global and local).
Some filter-based methods use global measures, such as the link betweenness-based strategy \cite{back1}. It defines network backbones formed only from links with the highest betweenness centrality. This method uses a threshold on the betweenness of links to preserve only those that exceed it. 
The link salience is another robust method based on a global measure \cite{back4}. This approach is based on the notion of the shortest-path tree that is summarizing the shortest paths from a reference node to the remainder of the network. It defines an average shortest-path tree matrix $S$, where each value $s_{ij}$ measures the number of shortest-path trees that the link $(i,j)$ is part of. Links with values higher than a defined threshold are kept, and they compose the backbone of the network. 

Other methods use local measures to extract the backbone. The k-core is a well-known measure used by J. Chalupa et al. \cite{back5,carmi2007model} as a hierarchical topology filter. For weighted networks, the link weight-based strategy \cite{back6} defines the $w$-skeleton or $w$-backbone obtained by discarding all the links with weights less than $w$. Arguing that the strength alone is not sufficient to capture the weighted structure of nodes, Serrano et al \cite{serrano} propose the disparity filter method. It is an improvement of the link weight-based method. This approach starts by normalizing the weights of each link $(i,j)$ attached to a given node $i$ by its strength (the sum of weights of all its connections). After that, a null model is considered, in which the weights of these links are supposed to be uniformly distributed. Then, the probability $\alpha_{ij}$ is computed. It indicates that the normalized weight of each link $(i,j)$ is compatible with the null model. The links with weights compatible with the uniform distribution are preserved with a significance level $\alpha$. Thus, edges verifying the condition ($\alpha_{ij} < \alpha$) should be kept in the backbone. The aim of this method is to highlight edges representing an important fraction of the local strength and weight magnitude of each given node.
In \cite{simmelian}, the authors introduce a local link filtering method exploiting Simmel's concept of membership in social groups. Experiments on Facebook interaction networks show that the Simmelian backbone is able to extract their main structure, making them easy to visualize and analyze.

The third type of filtering methods is based on a combination of local and global measures. In this context, Ronda J et al. \cite{hback} propose the $h$-Backbone approach which relies on three steps. Consider that the bridge measure of a link is equal to its betweenness divided by the total number of nodes of the network. In addition, $h_{b}$ represents the largest number such that there are $h_{b}$ links, each with bridge measure at least equal to $h_{b}$.
The first step consists of finding the links with a bridge measure value higher or equal to the $h$-bridge $h_{b}$. Let $h_{s}$ represents the largest number such that there are $h_{s}$ links, each with strength at least equal to $h_{s}$. The second step consists in searching for the links with a weight higher or equal to the $h$-strength $h_{s}$. In the last step, the edges selected from the two previous steps are merged to form the $h$-backbone. This approach selects relevant links connecting the network using the $h$-bridge, and high strength links located in the core of the network through the $h$-strength.

Backbone extraction processes dedicated to bipartite networks have been also developed \cite{neal1,neal2,fdsm,jo2020extracting}.  W. S. Jo \textit{et al.} \cite{jo2020extracting} proposed a new heuristic able to estimate the direction and strength of the hierarchical relationships between a pair of nodes in order to select the essential part of the networks. Experiments performed on two empirical datasets (the Gene Ontology network and a bipartite network of skills and individuals collected from LinkedIn) show that it manages to identify the most meaningful hierarchical relations from the bipartite networks. 
A R package \cite{backpackage} has been developed for bipartite networks in order to extract the backbone. It contains three methods: Hypergeometric Model (HM) \cite{neal1}, Stochastic Degree Sequence Model (SDSM) \cite{neal2} and Fixed Degree Sequence Model (FDSM) \cite{fdsm}.

Community structure is a feature commonly observed in complex networks. In real-world networks, nodes are usually found to be naturally arranged into various groups or communities. Numerous investigations have been conducted to model and to analyze the community structure of a network \cite{c1,lfrgunce,evans,tulu,jebablicomm,guimera2007classes,gunce4,cherifi2019community}. Yet, there is no unique interpretation of the community structure. It is regularly apprehended as the division of densely interconnected subgroups of nodes sparsely connected with nodes belonging to other modules. Communities can overlap. In an overlapping community structure, a node can belong to multiple communities, while it belongs to a single community in networks with a non-overlapping community structure. Palla et al. \cite{palla,membership} have shown that numerous real-world networks exhibit some nested and overlapping community structure. This behavior is quite natural in social networks. For instance, if we consider a network of actors, the actors may play in films of different genres (comedy, musical, adventure, action, etc). 
While there is a great deal of works exploiting the modular structure of real-world networks \cite{moi5,moi6,stra3,stra4,stra5,cbf,dcbf,pietro,bva,bridgeness,stra10,moi1,moi2}, and despite the ubiquity of this property, up to now, it is ignored in the filter-based backbone extraction approach. This article closes that gap. Indeed, in networks with an overlapping community structure, it is important to highlight the influence of the overlapping nodes and the hubs in the network structure and dynamics. 
On one hand, hubs are considered as the most influential nodes in the network. As they are the most connected, they can have a big influence on their community and also in bridging the communities. On the other hand, overlapping nodes are at the core of the network topology. Previous studies have shown that nodes belonging to the overlapping zones of the network are more tightly connected than nodes belonging to the non-overlapping zones \cite{leskovecover}. Therefore, the overlapping nodes can connect many nodes belonging to different parts of the network. They have, then, a big global influence on the network. To sum up, both hubs and overlapping nodes are important building blocks of the network that translate into a great local and global influence.

In this paper, we propose and investigate two backbone extraction methods that are based on these two types of nodes. The first method is called the overlapping nodes and hubs backbone. It is obtained by preserving the overlapping nodes and the hubs and removing the remaining nodes. All the edges with the lowest weights are then removed from this sub-network as long as the components stay connected. The second method is called the overlapping nodes ego backbone. In this case, the backbone is formed by considering the overlapping nodes and their one-step neighbors. Just as before, the sub-network is pruned in order to preserve the overlapping nodes and the neighbors with the highest weighted links without splitting the components. We expect that the overlapping nodes ego backbone is a good approximation of the overlapping nodes and hubs backbone. Indeed, in a previous study \cite{moi6}, it has been shown that the overlapping nodes are neighbors with a high number of the top connected nodes. Therefore, this backbone should preserve very relevant information from the original network without looking for the hubs that are supposed to belong to the overlapping nodes neighborhood. 
One can notice that unlike the vast majority of alternative filter-based methods, the proposed methods do not rely exclusively on links but also on nodes. Furthermore, both methods are local if the community structure is known.

A series of experiments are conducted on real-world weighted networks from different fields (social, co-appearance, collaboration, biological and technological) and various sizes. 
First of all, different evaluation measures (proportion of common nodes, rank-biased overlap, Pearson and Kendall Tau correlation) are used to measure the similarity and the correlation between the proposed backbone extraction methods. Then, they are compared to the Serrano backbone (called also the disparity filter backbone). Indeed, this filtering scheme is considered as one of the most effective local-based extraction methods.

\section*{Results and discussion}
In this section, results of the comparative evaluation of the backbone extraction methods are presented and analyzed. Samples of various real-world networks (social, co-appearance, biological, technological and collaboration networks) have been used in order to cover a wide range of situations. The overlapping community structure of the networks has been uncovered by the Speaker-Listener Label Propagation Algorithm (SLPA) \cite{slpa}. Results with a single community detection algorithm are reported. Indeed, previous results using alternative community detection algorithm have shown that the influence of the community detection algorithm is limited, and that the community structures uncovered are generally quite consistent\cite{moi6}.

In the first set of experiments, various evaluation criteria are used to measure the similarity between the two proposed backbones, and to compare them with the output of the disparity filter proposed by Serrano et al. \cite{serrano}. There are two main reasons for this choice. First, the disparity filter technique is based on a local measure. Thus, it belongs to the same category that the proposed methods. Second, it is considered as one of the most popular and effective alternative backbone extraction methods. To quantify the similarity between two backbones, the proportion of common nodes as well as the rank-biased overlap measures are computed (refer to the Materials and Methods section). Additionally, the statistical association between the set of nodes found in the backbones is measured by computing both Pearson and Kendall Tau correlation. 

In the second set of experiments, the effectiveness of the three extraction backbone methods is investigated. The average link weight, average node betweenness and weighted degree of the backbones are compared. The higher these values the better the backbone. Note that in all the experiments, the size of the backbone is limited to 30\% of the size of the original network by setting the parameter $s$ to 0.3. Accordingly, the parameter $\alpha$ is tuned in order to obtain a disparity filter backbone of the same size. 

\begin{table}[t!]
\centering
 \caption{The estimated parameters of the empirical networks under test. $N$ is the total number of nodes of the network. $A_{n}$ is the fraction of the common nodes between two sets of nodes of size $n$. $A_{t}$ is the proportion of the highly connected nodes of a given set of nodes, where $t$ is equal to 10\% of the top-ranked nodes of the network ($t=0.1*N$). Note that OE stands for the Overlapping nodes Ego backbone, OH stands for the Overlapping nodes and Hubs backbone while DF stands for the Disparity Filter backbone. Each reported value is the average of 10 SLPA simulation runs. The standard deviation is omitted from this table due to its very small value. It ranges between 0\% and 2\%.}
\label{t1}
\begin{tabular}{l|c|ccc|ccc}
\hline
\multicolumn{1}{c|}{\multirow{2}{*}{\textbf{Network}}} & \multirow{2}{*}{\textbf{$N$}} & \multicolumn{3}{c|}{\textbf{$A_{n}(\%)$}} & \multicolumn{3}{c}{\textbf{$A_{t}(\%)$}}                                        \\ \cline{3-8} 
\multicolumn{1}{c|}{}                                  &                               & OE - OH      & OE - DF     & OH - DF     & OE                        & OH                      & DF                        \\ \hline
Zachary’s karate club                                  & 34                            & 79           & 68.75       & 75          & 100                       & 100                     & 100                       \\
Intra-organisational                                   & 46                            & 100.0        & 84.61       & 84.61       & 100                       & 100                     & 50                        \\
Freeman’s EIES                                         & 48                            & 100.0        & 80.0        & 80.0        & 100                       & 100                     & 80                        \\
Train bombing                                          & 62                            & 100.0        & 94.73       & 94.73       & 100                       & 100                     & 100                       \\
Les Miserables                                         & 77                            & 91.66        & 86.11       & 88.88       & 100                       & 100                     & 93.33                     \\
Game of thrones                                        & 107                           & 100.0        & 68.75       & 68.75       & 100                       & 100                     & 90                        \\
C.elegans Neural                                       & 306                           & 96.38        & 64.04       & 64.04       & 100                       & 100                     & 75.86                     \\
Facebook-like Forum                                    & 899                           & 95.53        & 61.71       & 60.96       & 100                       & 100                     & 70.78                     \\
Facebook-like Social                                   & 1899                          & 100.0        & 75.04       & 75.04       & 100                       & 100                     & 91.53                     \\
US Power Grid                                          & 4941                          & 79.85        & 61.03       & 64.39       & \multicolumn{1}{l}{88.41} & \multicolumn{1}{l}{100} & \multicolumn{1}{l}{72.26} \\
Scientific Collaboration                               & \; 16726   \;                      &       91.34       &      57.83       &     59.06        &            90.28               &         97.05                & 68.19                          \\ \hline
\end{tabular}
\end{table}

\subsection*{Proportion of common nodes}
\begin{figure}[H]
\begin{center}
\includegraphics[width=18cm,height=18cm]{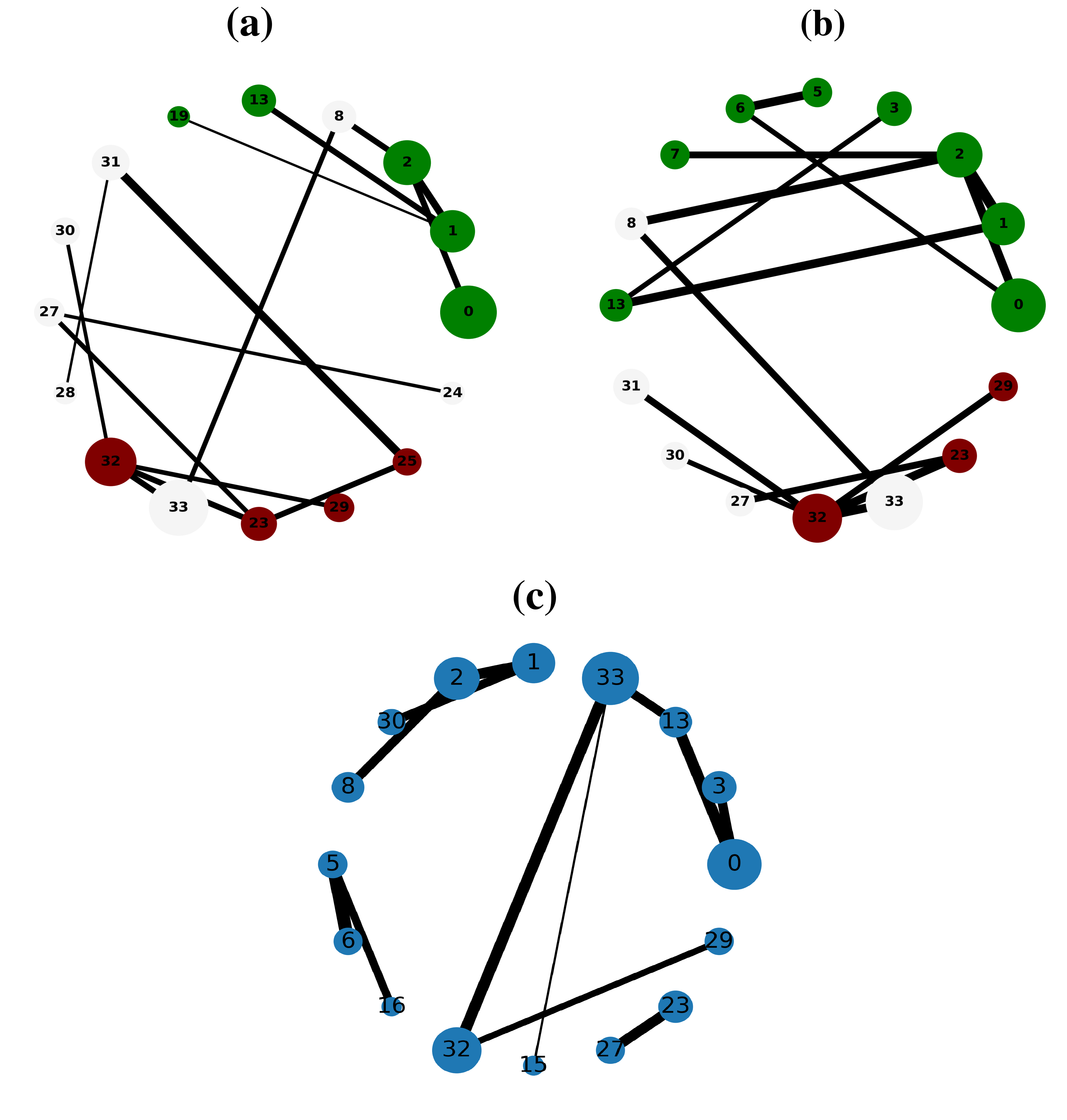}
\caption{The backbone extraction of different methods for the Karate club network. (a) overlapping nodes ego backbone, (b) overlapping nodes and hubs backbone and (c) Serrano backbone. Nodes are highlighted in different colors according to the community they belong to. Nodes with the same color belong to the same community while those in gray represent the overlapping nodes. The size of the nodes is proportional to their weighted degree, while the size of links is proportional to their weights. The community structure is revealed using the SLPA detection algorithm.    
\label{f3}
}
\end{center}
\end{figure}

\begin{figure}[H]
\begin{center}
\includegraphics[width=18cm,height=18cm]{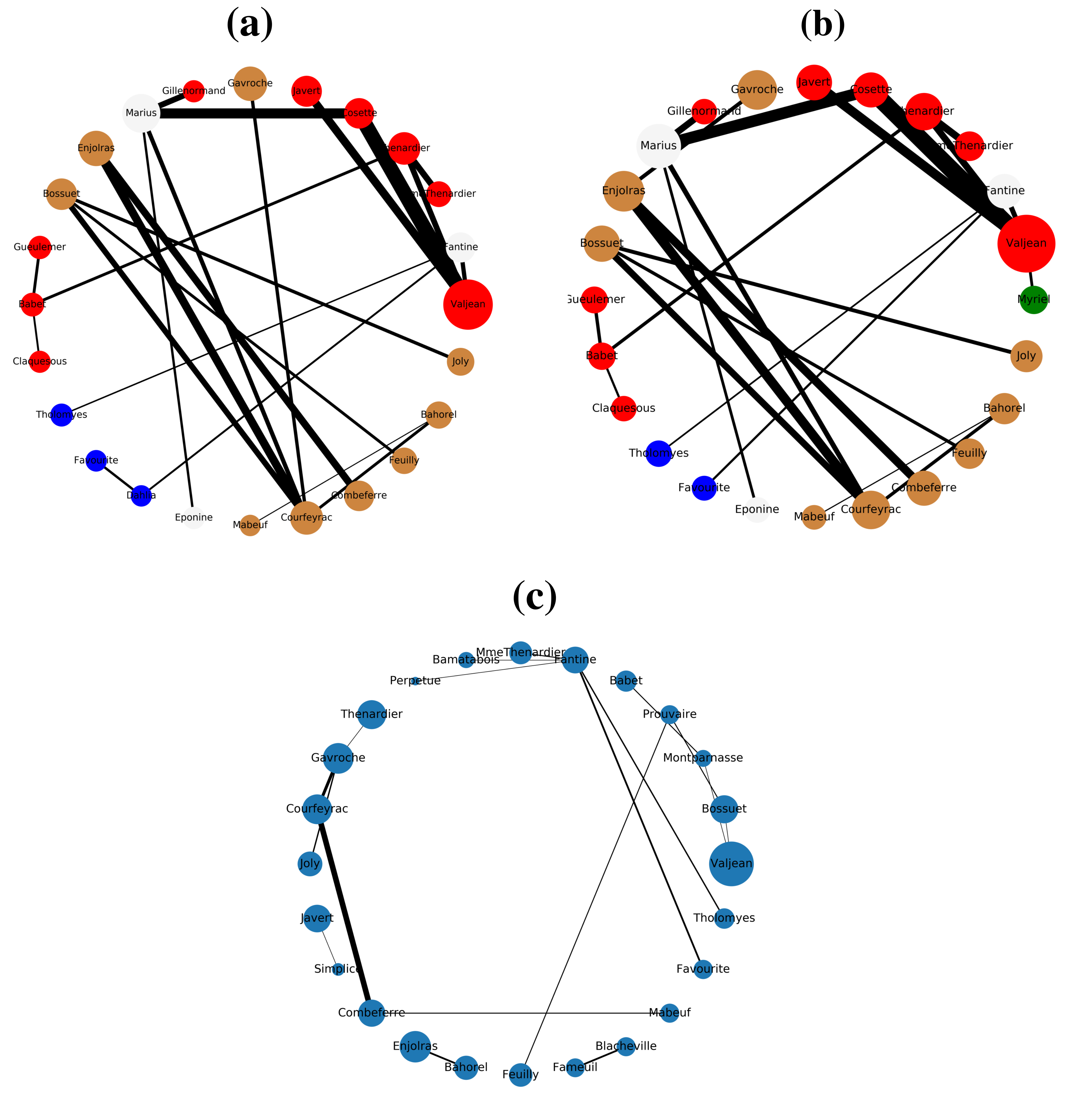}
\caption{The backbone extraction of different methods for Les Miserables network. (a) overlapping nodes ego backbone, (b) overlapping nodes and hubs backbone and (c) Serrano backbone. Nodes are highlighted in different colors according to the community they belong to. Nodes with the same color belong to the same community while those in gray represent the overlapping nodes. The size of the nodes is proportional to their weighted degree, while the size of links is proportional to their weights. The community structure is revealed using the SLPA detection algorithm.    
\label{f4}
}
\end{center}
\end{figure}

\begin{figure}[H]
\begin{center}
\includegraphics[width=18cm,height=18cm]{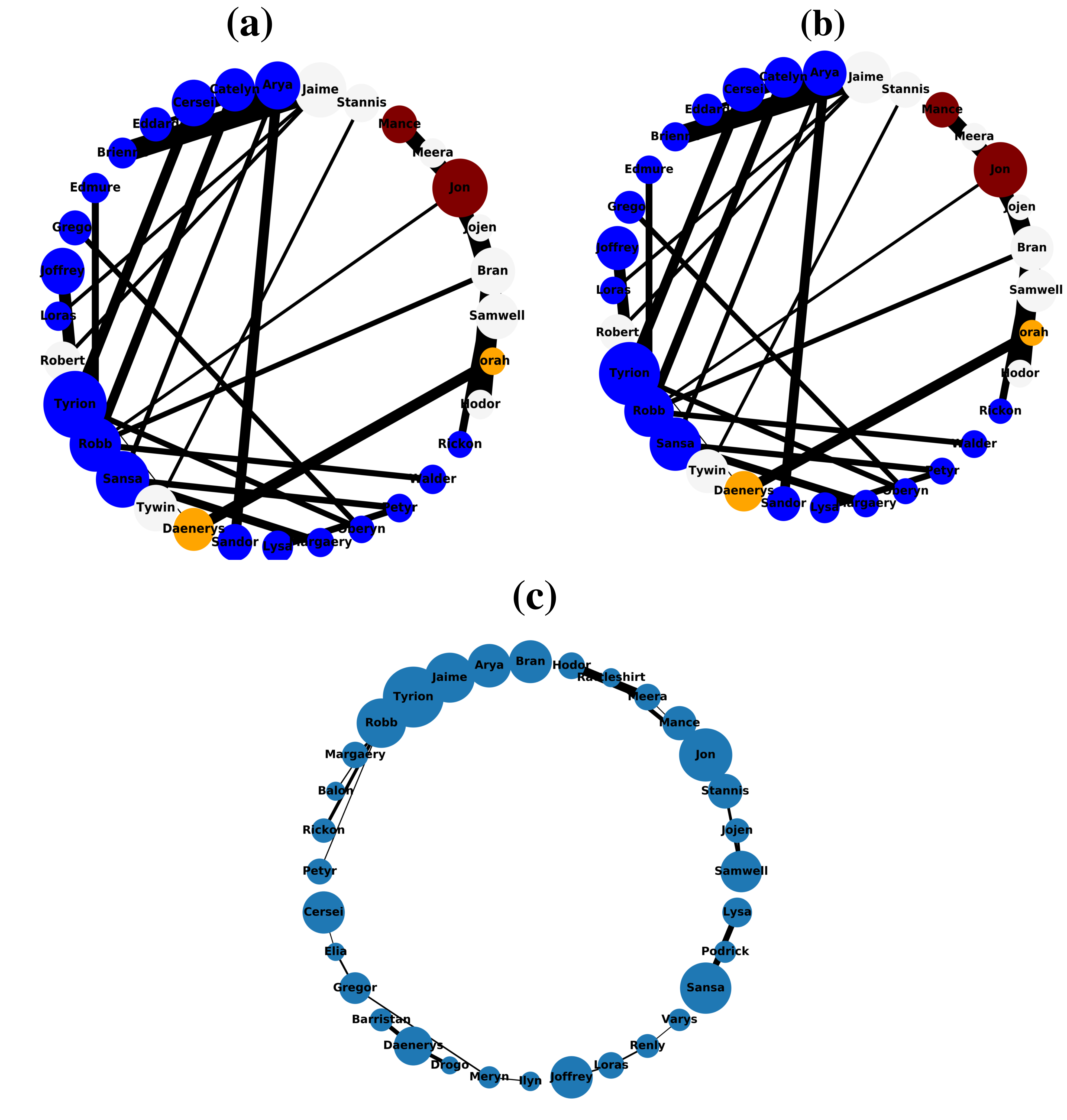}
\caption{The backbone extraction of different methods for the Game of thrones network. (a) overlapping nodes ego backbone, (b) overlapping nodes and hubs backbone and (c) Serrano backbone. Nodes are highlighted in different colors according to the community they belong to. Nodes with the same color belong to the same community while those in gray represent the overlapping nodes. The size of the nodes is proportional to their weighted degree, while the size of links is proportional to their weights. The community structure is revealed using the SLPA detection algorithm.    
\label{f7}
}
\end{center}
\end{figure}

\begin{figure}[H]
\begin{center}
\includegraphics[width=18cm,height=18cm]{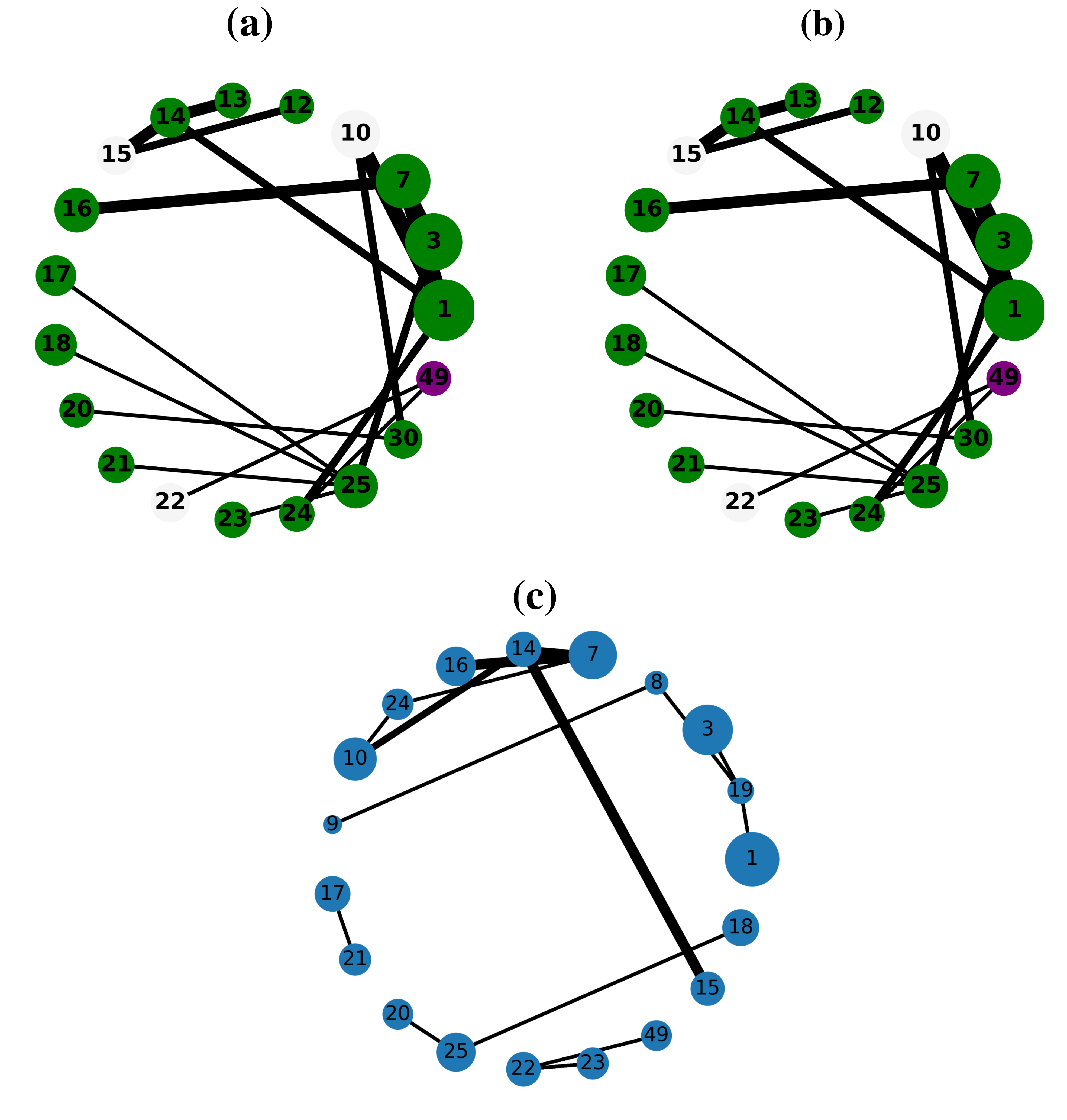}
\caption{The backbone extraction of different methods for the Train bombing network. (a) overlapping nodes ego backbone, (b) overlapping nodes and hubs backbone and (c) Serrano backbone. Nodes are highlighted in different colors according to the community they belong to. Nodes with the same color belong to the same community while those in gray represent the overlapping nodes. The size of the nodes is proportional to their weighted degree, while the size of links is proportional to their weights. The community structure is revealed using the SLPA detection algorithm.    
\label{f5}
}
\end{center}
\end{figure}

\begin{figure}[H]
\begin{center}
\includegraphics[width=18cm,height=18cm]{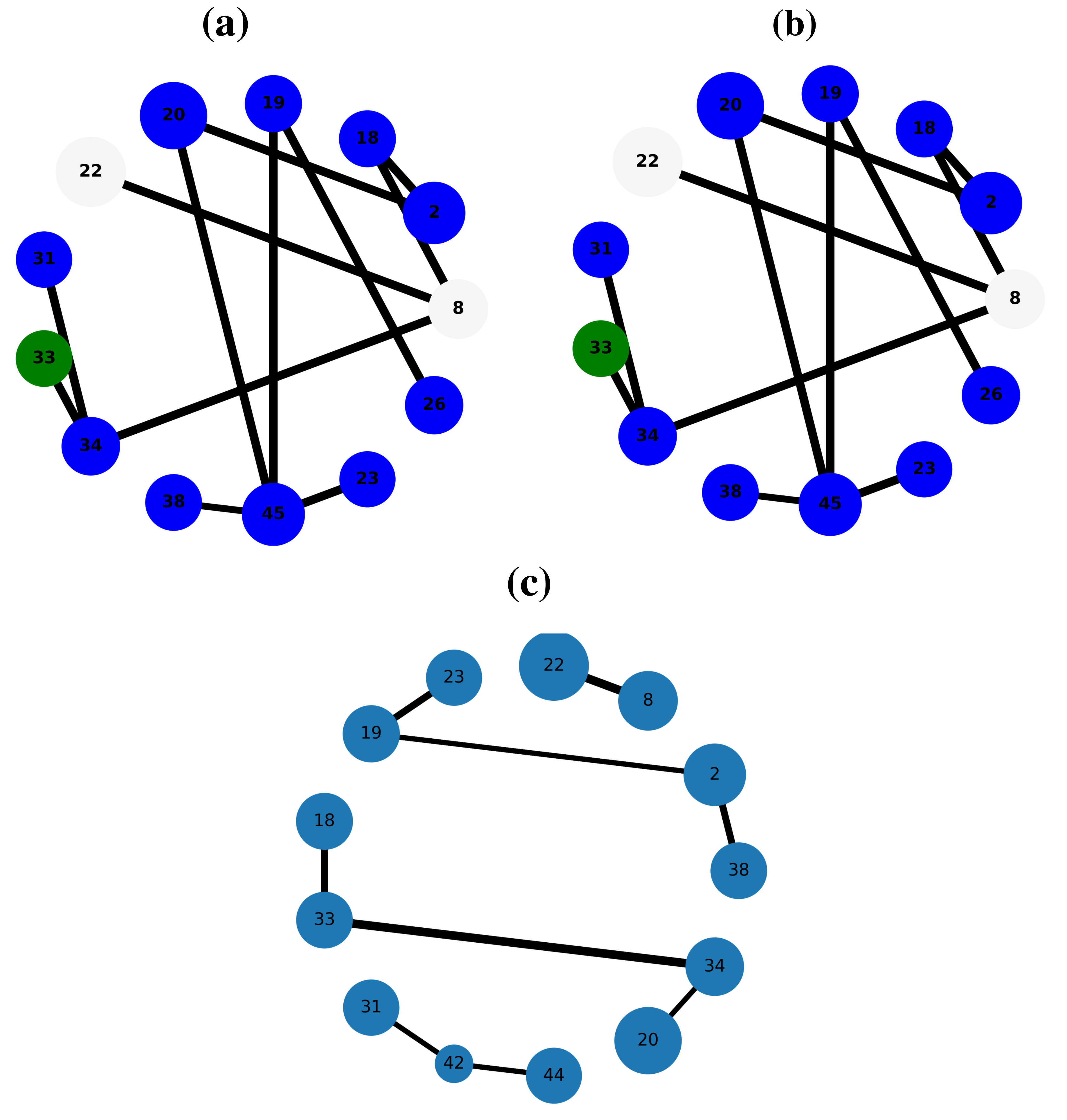}
\caption{The backbone extraction of different methods for the Intra-organisational network. (a) overlapping nodes ego backbone, (b) overlapping nodes and hubs backbone and (c) Serrano backbone. Nodes are highlighted in different colors according to the community they belong to. Nodes with the same color belong to the same community while those in gray represent the overlapping nodes. The size of the nodes is proportional to their weighted degree, while the size of links is proportional to their weights. The community structure is revealed using the SLPA detection algorithm.    
\label{f6}
}
\end{center}
\end{figure}

In this experiment, the proportion of common nodes extracted by the various backbone extraction methods is computed. Our goal is to check if they extract the same set of nodes or not. Let's first take a look at the extracted backbones in small size networks. In this case, the set of nodes can be compared visually. The \cref{f3,f4,f7,f5,f6} show the backbone extracted from Zachary's Karate Club, Les Miserables, Game of thrones, Train bombing and Intra-organizational networks respectively. In all the figures, the overlapping nodes ego backbone is represented in (a), the overlapping nodes and hubs backbone is represented in (b), while the disparity filter backbone is reported in (c). The color of nodes refers to the communities to which they belong. The overlapping nodes are colored in grey. The size of the nodes is proportional to their weighted degree \cite{wdegree} and the thickness of the links is proportional to their weights. These networks have a number of overlapping communities ranging from 2 to 3. One can see that, whatever the network considered, the overlapping nodes ego backbone and the overlapping nodes and hubs backbone share a vast majority of nodes.

Let's compare first the overlapping nodes ego backbones with the overlapping nodes and hubs backbones. In the Karate club network, 79\% of the nodes appear in both backbones. This number goes up to 91.6\% in Les Miserables network. The two backbones are identical for Game of thrones, Train bombing and Intra-organizational networks, as illustrated in  figures \ref{f7}, \ref{f5} and \ref{f6} respectively.

These results are consistent with the findings of a previous study \cite{moi6} showing that the overlapping nodes are neighbors of the highly connected nodes of the network. Therefore, the overlapping nodes ego backbones and the overlapping nodes and hubs backbones preserve just about the same set of nodes. In this situation, it is better to use the overlapping nodes ego backbone extractor as it needs less knowledge about the original network. The comparison with the disparity filter shows more differences. Indeed, the proportion of common nodes between the proposed backbones and the disparity filter backbones is relatively smaller for all the network under study. \cref{f3,f4,f7,f5,f6} illustrate this behavior. The main difference lies in the fact that the disparity filter concentrates on links while the proposed methods are based on nodes in their extraction process. In any case, in small networks, it appears clearly that the proposed backbone extractors manage to preserve more influential nodes and links from the original network as compared to the disparity filter algorithm.
Let's take Les Miserables network as an example, it is well known that Jean Valjean is a central character of Victor Hugo's novel. This ex-convict was running away for a great deal of time from the inspector Javert. Additionally, he is strongly attached to Cosette, his adopted daughter, and her future husband Marius who plays a major role in the french revolution. Mr and Mme Thenardier, the Parisian street child Gavroche, and Fantine, Cosette’s mother are also considered as very important characters of the novel. As a matter of fact, all these characters are selected by the proposed backbones. Furthermore, some main characters of this famous novel (e.g., Cosette and Marius) do not appear in the disparity filter backbone while some secondary characters are extracted by this algorithm. Some important links are also missed in the disparity filter backbone. For instance, the link between Valjean and Javert should not be removed from the backbone since these two characters have co-appeared in so many chapters of the novel. Game of thrones is another example where Catelyn Stark does not appear in the disparity filter backbone despite its main role in this saga. Therefore, the backbones obtained by the proposed approaches preserve almost all high-connectivity nodes and essential connections.

\autoref{t1} reports the results for all the networks under test. The overlap between the overlapping nodes ego and the overlapping nodes and hubs backbones is ranging from 79\% to 100\%. It shows that the hubs are tightly connected to the overlapping nodes that are located in the core of the communities. 

Furthermore, we also check if the top-ranked nodes of the original network are preserved in the backbones. The set of the 10\% top-ranked nodes according to their weighted degree is computed in the original network. It is compared to the set of top-ranked nodes of the same size extracted from the backbones. Results show that both sets are identical most of the time for the proposed backbones. This remark is valid for all the empirical networks under evaluation except US power grid and scientific collaboration that exhibit an overlap higher than 90\%. As expected, the ability of the overlapping nodes and hubs backbone to preserve the highly connected nodes is slightly higher.

Both measures are also computed to assess the similarity between the proposed backbones and the disparity filter backbone. Results show that there is a relatively smaller overlap between the proposed backbones and the disparity filter backbone. Indeed, the overlap between the disparity filter and the overlapping nodes ego backbones ranges from 57.83\% to 94.73\% while it ranges from 59.06\% to 94.73\%  for the overlapping nodes and hubs backbone. 
Comparisons with the set of the 10\% top-ranked nodes of the original networks show that it preserves a smaller proportion of the top-ranked nodes. Indeed, the disparity filter backbone focuses on links rather than on nodes. That is the main difference between the proposed backbones and the disparity filter backbone that explain its lower ability to preserve hubs.

\begin{table}[t!]

 \caption{The estimated values of the rank-biased overlap $r$ between the sets of nodes of two backbones OE stands for the overlapping nodes ego backbone, OH stands for the overlapping nodes and hubs backbone while DF stands for the disparity filter backbone. The parameter $p$ defines the weights of the elements of the two sets. Equal weight is assigned to all the elements of both sets when $p$ has a high value (close to 1). Each value is the average of 10 SLPA simulation runs. The standard deviation is excluded from this table due to its very small value (it is around $10^{-3}$).}
\label{t2}
\footnotesize\setlength{\tabcolsep}{4.5pt}

\begin{tabular}{c|ccccc|ccccc|ccccc}
\hline
\multirow{3}{*}{\textbf{Network}} & \multicolumn{15}{c}{\textbf{$r$}}                                                                                                                                                                                                                              \\ \cline{2-16} 
                                  & \multicolumn{5}{c|}{\textbf{OE - OH}}                                               & \multicolumn{5}{c|}{\textbf{OE - DF}}                                               & \multicolumn{5}{c}{\textbf{OH - DF}}                                               \\ \cline{2-16} 
                                  & \textbf{p=0.5} & \textbf{p=0.7} & \textbf{p=0.8} & \textbf{p=0.9} & \textbf{p=0.98} & \textbf{p=0.5} & \textbf{p=0.7} & \textbf{p=0.8} & \textbf{p=0.9} & \textbf{p=0.98} & \textbf{p=0.5} & \textbf{p=0.7} & \textbf{p=0.8} & \textbf{p=0.9} & \textbf{p=0.98} \\ \hline
Zachary’s karate club                                 & 1          & 0.98          & 0.96          & 0.93          & 0.89           & 0.94          & 0.93          & 0.91          & 0.85          & 0.74           & 0.95          & 0.94          & 0.94          & 0.91          & 0.85           \\
Intra-organisational                             & 1              & 1              & 1              & 1              & 0.99           & 0.85          & 0.77          & 0.77          & 0.74          & 0.74           & 0.85          & 0.77          & 0.77          & 0.74          & 0.74           \\
Freeman’s EIES                           & 1              & 1              & 1              & 1              & 0.99           & 0.94          & 0.91           & 0.89          & 0.87          & 0.85            & 0.94          & 0.91           & 0.89          & 0.87          & 0.85            \\
Train bombing                                & 1              & 1              & 1              & 1              & 0.99           & 0.95          & 0.95          & 0.95          & 0.94          & 0.91            & 0.95          & 0.95          & 0.95          & 0.94          & 0.91            \\
Les Miserables                      & 1          & 1          & 0.99          & 0.98          & 0.94            & 0.95          & 0.95          & 0.94          & 0.93          & 0.88           & 0.95          & 0.95          & 0.94          & 0.93          & 0.88           \\
Game of thrones                        & 1              & 1          & 1          & 1          & 1           & 0.95          & 0.94          & 0.92          & 0.86          & 0.71           & 0.95          & 0.94          & 0.92          & 0.86          & 0.71           \\
C.elegans Neural                         & 0.99          & 0.99          & 0.99          & 0.99          & 0.98           & 0.95          & 0.94           & 0.92          & 0.87          & 0.65           & 0.95          & 0.94           & 0.92          & 0.87          & 0.65           \\
Facebook-like Forum                   & 1              & 0.99          & 0.99          & 0.99          & 0.99           & 0.95         & 0.94          & 0.94          & 0.9          & 0.77           & 0.95          & 0.94          & 0.94          & 0.9          & 0.77           \\
Facebook-like Social               & 1              & 1          & 1          & 1          & 1           & 0.95          & 0.94          & 0.94          & 0.92          & 0.91           & 0.95          & 0.94          & 0.94          & 0.92          & 0.91           \\
US Power Grid                        & 0.99          & 0.99          & 0.98          & 0.97          & 0.97           & 0.94          & 0.91           & 0.89          & 0.87          & 0.87           & 0.94          & 0.91           & 0.91          & 0.89          & 0.88           \\
Scientific Collaboration           &        1        &       0.99         &     0.99           &             0.98    &       0.97         &         0.96         &      0.95        &       0.95                   &     0.88  &        0.84         &       0.96         &      0.96          &      0.95          &   0.92             &        0.9         \\ \hline
\end{tabular}
\end{table}

\subsection*{Rank-biased overlap}
 The rank-biased overlap (RBO) measures the similarity between two sets. It quantifies the proportion of common nodes found in two sets while incrementally increasing their depths. Its value is equal to 1 when the two sets are similar, while it is equal to 0 when the sets are totally disjoint. Different weights are assigned to the elements of both sets while computing the RBO. The weights of the elements are set using the parameter $p$ (refer to the Materials and Methods section). More weight is given to the comparison of the top elements of the two sets if this parameter has a small value. 

At first, the RBO is computed between the set of nodes of the two proposed backbones. Note that the sets of nodes are sorted in decreasing order according to their weighted degree. The experimental results are reported in \autoref{t2}. One can see that the overlap between the two sets is quite high when nearly all the elements have the same weight ($p=0.98$). It ranges between 88\% and 99\%. The values of the RBO gets even higher when more importance is assigned to the top-ranked elements ($p=0.5$). For all the networks, it is always close to 100\%. Thus, these results confirm and reinforce the previous observations. This is more visible in the small networks illustrated in \cref{f3,f4,f7,f5,f6}, where the highly connected nodes are selected by both backbones. Secondly, \autoref{t2} reports also the values of the same metric computed between the proposed backbones and the disparity filter backbone. It can be noticed from the results that the RBO exhibit high values for the various values of the parameter $p$. However, they are relatively smaller than the ones obtained when this measure is computed between the two proposed backbones. Indeed, the disparity filter backbone sometimes includes some peripheral nodes which do not have a big influence in the network. These nodes are not selected by the overlapping nodes ego and the overlapping nodes and hubs backbone that extract nodes located in the core of the network. To summarize, there is a great similarity between the sets of nodes found in the two proposed backbones. Moreover, the top-ranked nodes are always preserved in these backbones. Yet, there is relatively less similarity between these backbones and the disparity filter backbone. The latter one sometimes selects less relevant nodes.

\begin{table}[t!]
\centering
 \caption{The Correlation between the sets of nodes of two backbones. For shortness, OE stands for the overlapping nodes ego backbone, OH stands for the overlapping nodes and hubs backbone while DF stands for the disparity filter backbone. $\rho$ represents the Pearson correlation while $\rho_{\tau}$ represents the Kendall Tau correlation. Each value is the average of 10 SLPA simulation runs. The standard deviation is omitted from this table due to its very small value (it is around $10^{-3}$).}
\label{t3}
\begin{tabular}{l|ccc|ccc}
\hline
\multicolumn{1}{c|}{\multirow{2}{*}{\textbf{Network}}} & \multicolumn{3}{c|}{\textbf{$\rho$}}                   & \multicolumn{3}{c}{\textbf{$\rho_{\tau}$}}             \\ \cline{2-7} 
\multicolumn{1}{c|}{}                                  & \textbf{OE - OH} & \textbf{OE - DF} & \textbf{OH - DF} & \textbf{OE - OH} & \textbf{OE - DF} & \textbf{OH - DF} \\ \hline
Zachary’s karate club                                   & 0.99            & 0.97             & 0.98            & 0.99            & 0.95            & 0.96            \\
Intra-organisational                                   & 1                & 0.89            & 0.89            & 0.99            & 0.99            & 0.99            \\
Freeman’s EIES                                          & 1                & 0.58            & 0.58            & 1                & 0.98            & 0.98            \\
Train bombing                                          & 1                & 0.81            & 0.81            & 1                & 0.98            & 0.98            \\
Les Miserables                                         & 0.99            & 0.79            & 0.8            & 0.99           & 0.97            & 0.98            \\
Game of thrones                                       & 1                & 0.91            & 0.91            & 1                & 0.93                & 0.93                \\
C.elegans Neural                                        & 0.99            & 0.8            & 0.81            & 0.99            & 0.98            & 0.98            \\
Facebook-like Forum                                     & 1                & 0.91            & 0.91            & 1                & 0.98            & 0.98            \\
Facebook-like Social                                   & 1                & 0.91            & 0.91            & 1                & 0.99            & 0.99            \\
US Power Grid                                          & 0.97            & 0.87            & 0.88            & 0.91            & 0.86            & 0.87            \\
Scientific Collaboration                                &        0.99          &       0.86           &                 0.87  &          0.99        &       0.98           &     0.98            \\ \hline
\end{tabular}
\end{table}

\subsection*{Correlation}
Here, the correlation between the set of nodes found in the different backbones is computed. Two correlation measures are used: Pearson and Kendall Tau correlation. The degree of relationship between the sets of nodes is estimated using the Pearson correlation given in \autoref{eq1}. It is computed using the weighted degrees of the sets of nodes sorted in decreasing order. The Kendall Tau correlation is also used to assess the statistical associations based on the ranks of these sets. We compute the ranks of the sets according to the weighted degree of nodes. A different rank is associated with each degree. Both correlation metrics have values ranging between -1 and 1. The sets of nodes have a perfect positive correlation when their values are equal to 1, while they have perfect negative correlation when the values are equal to -1.

\autoref{t3} reports the empirical results of Pearson and Kendall Tau correlation for all the real-world networks. It can be noticed that the values of both correlation measures computed between the sets of nodes extracted by the proposed backbones are very high. They are very close to 1. This is due to the very high fraction of nodes they have in common. Thus, both sets have about the same degrees and ranks. These results confirm that there is a very strong relationship between the sets of nodes of the two backbones. Additionally, the correlation measures have relatively lower values between the two backbones and the disparity filter backbone. The values are still high (ranging between 0.8 and 0.9), but they are lower than the ones obtained between the proposed backbones. These results are in concordance with previous empirical findings. Indeed, in some networks, the disparity filter backbone includes some different nodes with a very small weighted degree. Therefore, there is a relatively weaker monotonic relationship between the proposed backbones and this alternative one.

\begin{table}[t!]
\centering  
 \caption{The estimated values of the average weighted degree, betweenness and the link weight of a backbone. For shortness, OE stands for the overlapping nodes ego backbone, OH stands for the overlapping nodes and hubs backbone while DF stands for the disparity filter backbone. $<k>$ represents the average weighted degree. $<\beta>$ is the average node betweenness while $<w>$ average link weight. Each value is the average of 10 SLPA simulation runs. The standard deviation is omitted from this table due to its very small value. It is around $10^{-3}$ for the average betweenness, and it ranges between 0.1 and 0.3 for the average weighted degree and link weight.}
\label{t4}
\begin{tabular}{l|ccc|ccc|ccc}
\hline
\multicolumn{1}{c|}{\multirow{2}{*}{\textbf{Network}}} & \multicolumn{3}{c|}{\textbf{$<\beta>$}} & \multicolumn{3}{c|}{\textbf{$<k>$}}      & \multicolumn{3}{c}{\textbf{$<w>$}}       \\ \cline{2-10} 
\multicolumn{1}{c|}{}                                  & \textbf{OE}      & \textbf{OH}      & \textbf{DF}     & \textbf{OE} & \textbf{OH} & \textbf{DF} & \textbf{OE} & \textbf{OH} & \textbf{DF} \\ \hline
Zachary’s karate club                                  & 0.088            & 0.09             & 0.079           & 11.89      & 12.71      & 11.21      & 3.31       & 3.46       & 3.15        \\
Intra-organisational                                   & 0.028            & 0.028            & 0.013           & 49.42      & 49.42      & 39.35      & 2.31       & 2.31       & 2.18        \\
Freeman’s EIES                                         & 0.014            & 0.014            & 0.011           & 43.84       & 43.84       & 38.81      & 2.53       & 2.53       & 2.14       \\
Train bombing                                          & 0.077            & 0.077            & 0.047           & 16.77      & 16.77      & 11.87      & 1.38        & 1.38        & 1.23       \\
Les Miserables                                         & 0.057            & 0.064            & 0.035           & 25.36      & 25.58      & 16.13      & 4.76       & 4.89       & 3.74       \\
Game of thrones                                        & 0.053            & 0.053            & 0.034           & 51.47      & 51.47      & 37.22      & 16.58      & 16.58      & 14.98      \\
C.elegans Neural                                       & 0.012            & 0.012            & 0.009           & 52.35       & 52.66      & 39.91      & 4.95       & 5.32       & 4.91       \\
Facebook-like Forum                                    & 0.005            & 0.005            & 0.004           & 79.77      & 79.77      & 56.93      & 6.99        & 6.99        & 5.61       \\
Facebook-like Social                                   & 0.003            & 0.003            & 0.002           & 638.79     & 638.79     & 463.24     & 356.19     & 356.19     & 313.51     \\
US Power Grid                                          & 0.008            & 0.009            & 0.005           & 4.3       & 4.67       & 3.51       & 50.02      & 53.59      & 49.55      \\
Scientific Collaboration                               &        0.008          &       0.008           &                0.006  &      16.19       &      16.53       &    11.98         &      49.94       &     49.97        &      48.52       \\ \hline
\end{tabular}
\end{table}

\subsection*{Effectiveness}
In this subsection, the performance of the three backbone extraction methods is compared by measuring the average betweenness, the weighted degree and the average link weight. \autoref{t4} reports the results for all networks under test.


Let's look first at the average betweenness. The betweenness measures the extent to which a node lies on paths between all the other nodes of the network. The average betweenness indicates how much information can pass through the nodes of the backbone. The greater the value, the better the backbone's ability to disseminate information. It can clearly notice from \autoref{t4} that both overlapping nodes ego and overlapping nodes and hubs backbone exhibit a very close average betweenness for all the networks. This is because both backbones select nearly the same set of nodes. However, the values of the average betweenness of both backbones are higher than the ones computed in the disparity filter backbone. This implies that the nodes extracted by the proposed backbones act as a better information gateway of the original network as compared to the disparity filter backbone. 

Let's now turn to the average weighted degree of the set of nodes selected by the different backbones. This measure reflects the connectedness of the nodes of a given backbone. The experimental results reported in \autoref{t4} corroborate the conclusions made with the average betweenness. Indeed, the proposed backbones display the same performance. For all the networks, they have an equal or a very close average weighted degree value. Yet, they outperform the disparity filter backbone in terms of the connectedness. This confirms our observations made in the small networks illustrated in \cref{f3,f4,f7,f5,f6}. Taking a direct look at these networks, it can be noticed that some highly connected nodes are missed by the disparity filter backbone. For instance, Marius and Cosette in Les Miserables network do not appear in this backbone, while other secondary characters are included. This explains its lower node connectedness as compared to the proposed backbones.

Finally, let's compare the average link weight. A higher value of this measure demonstrates that the picked links are quite relevant. Experimental results reported in \autoref{t4} show that the average link weight of the proposed backbones exhibits approximately the same values. It is slightly higher for the overlapping nodes and hubs backbone. The two backbones have, however, a higher average link weight as compared to the disparity filter backbone. Indeed, this backbone has fewer connections between its nodes. Some very relevant connections are missed in the disparity filter backbone. As mentioned before in the previous subsection, the link between Jean Valjean and Javert in Les Miserables network is excluded from this backbone. So, the two proposed backbones select more relevant connections as compared to the disparity filter backbone. Furthermore, another disadvantage of the disparity filter backbone is that it is divided into many components while the proposed backbones are composed of a single connected component. Therefore, the topology of the sub-networks generated by the proposed approach preserves more the shape of the spine of the original network. 

To conclude, both overlapping nodes ego and overlapping nodes and hubs backbone display similar performance in terms of information gateway, connectedness and link strength. Their performance is higher than the disparity filter backbone for all the tested empirical networks. These results allow to shed more light on the overlapping nodes ego backbone. Indeed, this backbone has the same performance as the overlapping nodes and hubs backbone. Moreover, it requires only the local information about the overlapping nodes. Therefore, it can be considered as the most appropriate extraction method in large-scale networks.


\subsection*{Influence of the parameters setting}
The outputs from the proposed backbone extractors depend on the size of the overlapping nodes set $m$, the size of their neighbors set $k$, and the size of hubs set $t$. These parameters depend essentially on the revealed community structure. Indeed, the size of the set of the neighborhood of overlapping nodes is directly defined after identifying all the overlapping nodes of the network. Furthermore, the size of the set of hubs has the same size as the set of neighbors of the overlapping nodes. Rather than reporting results on the influence of the community detection algorithm on these parameters, we recall the main findings. For more details, the reader can refer to \cite{moi6}. In this previous study, results of extensive experiments using three community detection algorithms (Demon, an Improved version of LFM, SLPA) on a number of empirical networks of various origin are reported. They show that generally the size and the content of these sets are very stable and quite comparable. However, sometimes the size of the overlapping nodes set can be relatively large. In such cases, the values of the parameters $k$ and $t$ are also high. Therefore, the backbones have a relatively large size as well. This is the reason why the parameter $s$ controlling the size of the backbones is introduced. It limits the size of the backbone in cases where the default extracted backbones have a too large size. Top-ranking nodes are privileged in such situations.

\section*{Conclusion}
Understanding the properties and the topological structure of large-scale networks has become an increasingly challenging issue. Therefore, finding a way to extract the truly relevant nodes and connections in order to obtain a reduced and meaningful representation of these networks is a must. In this work, two local methods to extract backbones in weighted networks based on overlapping nodes are proposed. The overlapping nodes ego backbone is formed only from the set of overlapping nodes with their neighbors. The second backbone is formed with the set of overlapping nodes and the highly connected nodes of the network. It is called the overlapping nodes and hubs backbone. The links with the lowest weights are also removed as long as a single connected component is preserved. Both backbones extraction methods are tested on real-world networks selected from various fields. At first, the sets of nodes of the backbones are compared using similarity and correlation measures (the proportion of common nodes, rank-biased overlap, Pearson and Kendall Tau correlation). The first two measures show that there is a very high overlap between these backbones while the third and fourth measures show that there is a very high correlation between the sets of nodes. This confirms that the extracted backbone networks are almost identical. Additionally, the proposed backbone extractors are also compared with the disparity filter using the same measures. It is one of the most effective local extraction methods. Results show that there is a relatively smaller overlap between the backbones uncovered by the proposed methods and the disparity filter backbone. This is also confirmed by the correlation which displays smaller values. 
Furthermore, we also compare the effectiveness of the three backbones. To do so, the average weighted degree, betweenness and average link weight are computed. Results show that the overlapping nodes ego backbone is as effective as the overlapping nodes and hubs backbone in terms of information gateway, connectedness and link strength. Furthermore, their performance is higher than the disparity filter backbone. Therefore, the proposed backbones are very effective at preserving the more relevant nodes and connections.

\section*{Materials and Methods}\label{l1}

\subsection*{Extracting the network backbone}
In this study, the backbone of weighted networks is extracted using two different methods. The first one called "overlapping nodes ego backbone" consists of defining the set of overlapping nodes and their neighbors as the network backbone. The second method called the "overlapping nodes and hubs backbone" is based on selecting the overlapping nodes and the hubs of the network to form the network backbone. These two models are detailed in the next two paragraphs.

\subsubsection*{Overlapping nodes ego backbone} 
Let's consider a network $G(V,E)$, where $V = \{v_{1}, v_{2}, ...., v_{n}\}$ and $E = \{(v_{i}, v_{j}) \setminus v_{i}, v_{j} \in V \}$ denotes the set of nodes and edges respectively. The algorithm used to extract the overlapping nodes ego backbone is as follows:\\

\noindent \textbf{Step 1:} If not known, the community structure of the network using a given overlapping community detection algorithm is uncovered.\\

\begin{figure}[H]
\begin{center}
\includegraphics[width=18cm,height=18cm]{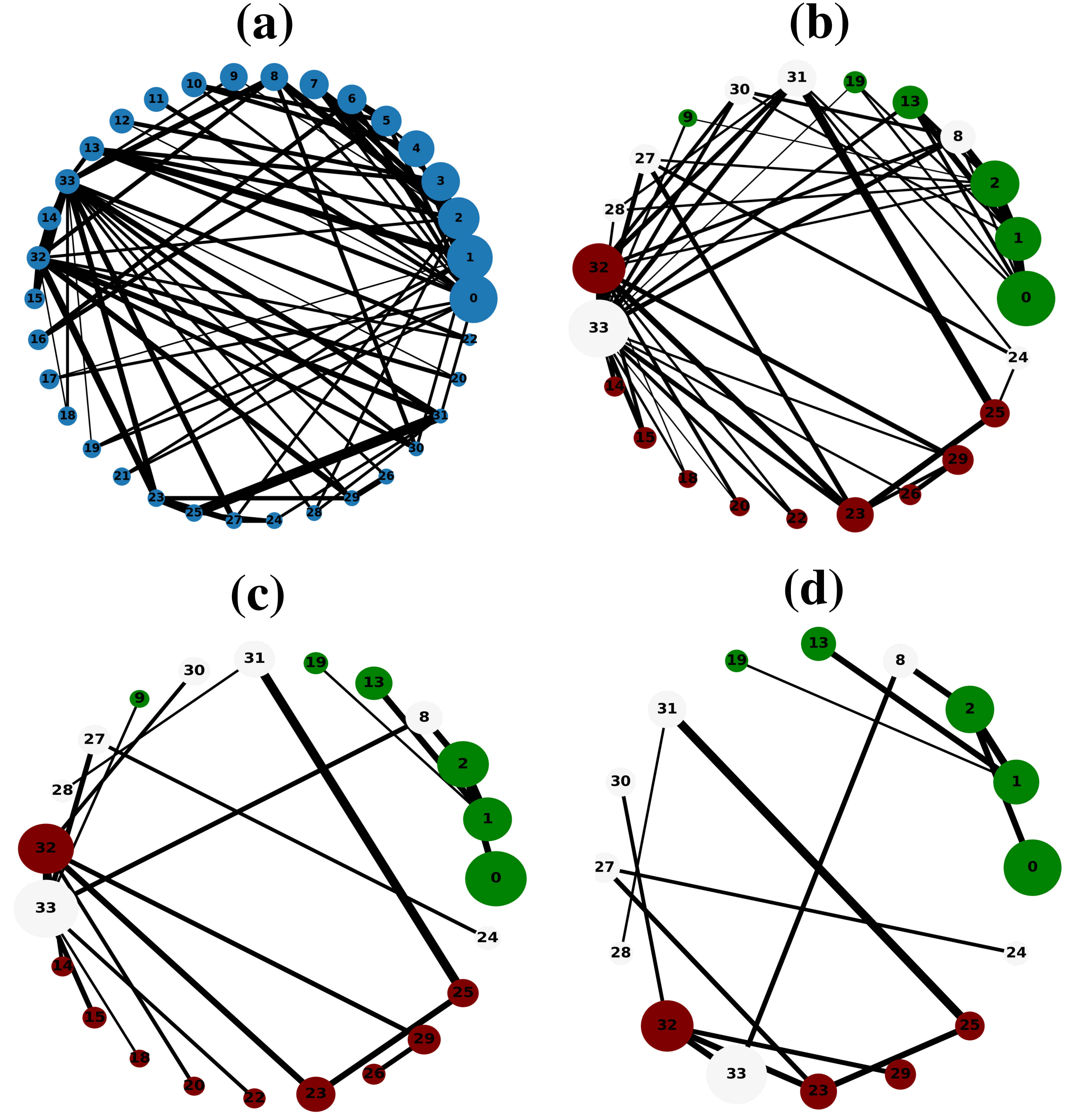}
   \centering
\caption{The general steps to extract the overlapping nodes ego backbone for a toy example (a). (b) illustrates the formation of the sub-network. All the links with low weights are removed in (c). (d) illustrates the sub-network with the parameter $s$ set to $0.4$. Nodes are highlighted in different colors according to the community they belong to. Nodes with the same color belong to the same community while those in gray represent the overlapping nodes. The size of the nodes is proportional to their weighted degree, while the size of links is proportional to their weights. The community structure is revealed using the SLPA detection algorithm.  
\label{f1}
}
\end{center}
\end{figure}

\noindent \textbf{Step 2:} Form the sub-network made of the overlapping nodes and their neighbors. The set of overlapping nodes is defined as  $V_{o} = \{v_{1}^{o}, v_{2}^{o}, ...., v_{m}^{o}\} \subset V$ where $m$ is the number of the overlapping nodes. The set of their neighbors is denoted by $V_{no} = \{v_{1}^{no}, v_{2}^{no}, ...., v_{k}^{no}\} \subset V$ where $k$ is the size of the neighborhood of the overlapping nodes. The overlapping nodes ego network is formed from the union of the set of overlapping nodes and the set of their neighbors. It is obtained by removing all nodes that do not belong to one of the two sets. The overlapping nodes ego network is denoted $M(V_{q},E_{q})$, where $V_{q}= V_{o} \cup V_{no}$ and $E_{q}=\{(v_{i}^{q}, v_{j}^{q}) \setminus v_{i}, v_{j} \in V_{q}\}$ are respectively its set of nodes and edges. \autoref{f1} (b) illustrates the formation of this sub-network for the toy example shown in \autoref{f1} (a).\\

\noindent \textbf{Step 3:} Remove the edges with the lowest weights from the overlapping ego sub-network. To do so, edges are sorted in decreasing order according to their weights. Then, edges with the lowest weights are removed as long as the sub-network remains connected. \autoref{f1} (c) illustrates this step, where all the links with the lowest weights are removed.\\

\noindent \textbf{Step 4:} Control the size of the overlapping ego backbone with a parameter $s$.  This parameter allows to preserve only the top-ranked nodes of this backbone. To this end, all the nodes of the overlapping nodes ego backbone are sorted in decreasing order according to the weighted degree centrality \cite{wdegree}. After that, according to the value of this threshold parameter, the nodes with low degrees are removed from the network. \autoref{f1} (d) illustrates the obtained overlapping nodes ego backbone by setting the parameter $s$ to $0.4$. In this case, the backbone represents $40\%$ of the nodes of the original network. 

\subsubsection*{Overlapping nodes and hubs backbone} 
Let’s consider a network $G(V,E)$, where $V = \{v_{1}, v_{2}, ...., v_{n}\}$ and $E = \{(v_{i}, v_{j}) \setminus v_{i}, v_{j} \in V \}$ denotes the set of nodes and edges respectively. To extract the overlapping nodes and hubs backbone, we follow the next steps:\\

\noindent \textbf{Step 1:} If not known, the community structure of the network using a given overlapping community detection algorithm is uncovered.\\

\begin{figure}[H]
\begin{center}
\includegraphics[width=18cm,height=18cm]{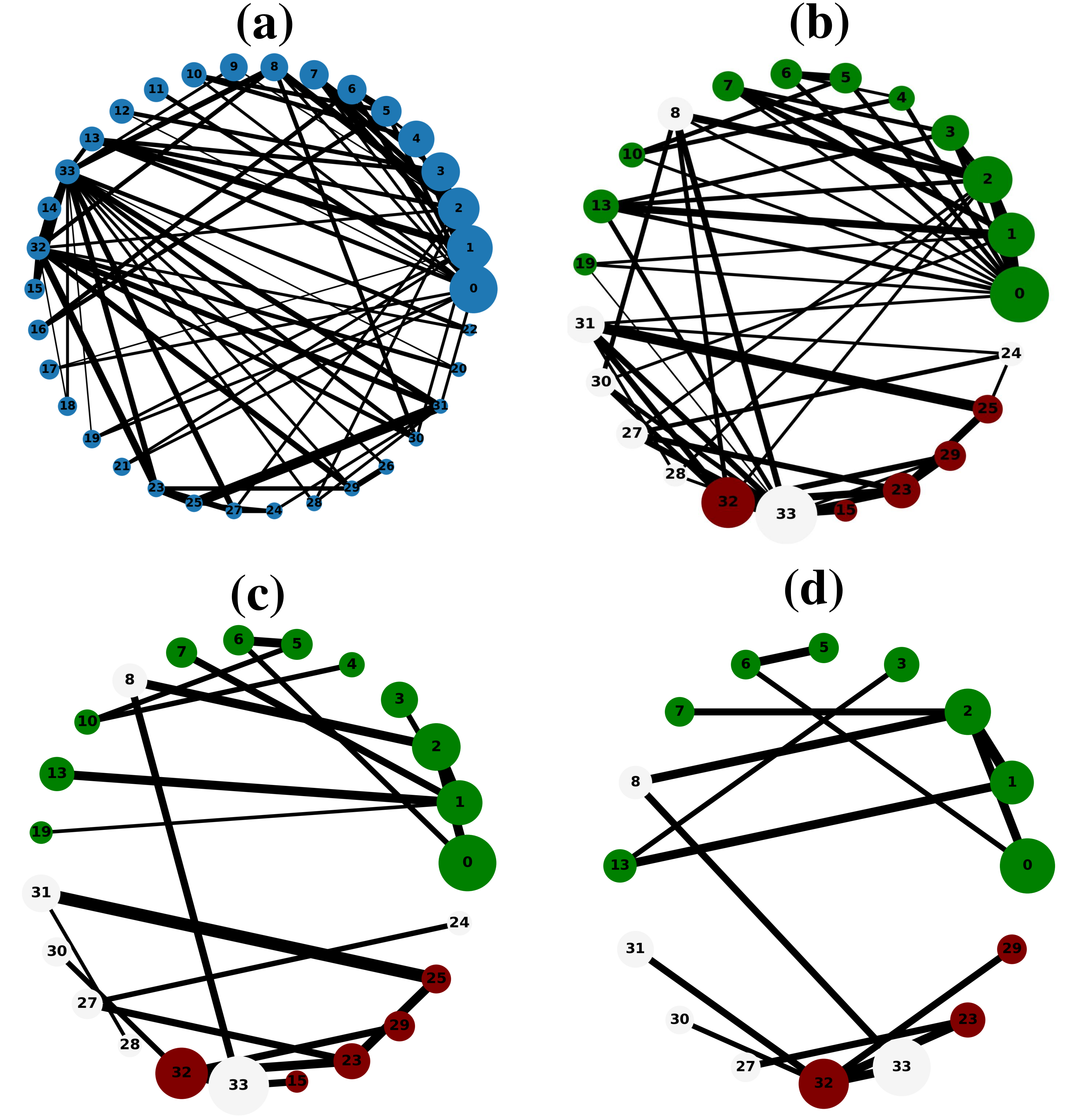}
   \centering
\caption{The general steps to extract the overlapping nodes and hubs backbone for a toy example (a). (b) illustrates the formation of the sub-network. All the links with low weights are discarded in (c). (d) shows the sub-network with the size threshold parameter $s$ set to $0.4$.  Nodes are highlighted in different colors according to the community they belong to. Nodes with the same color belong to the same community while those in gray represent the overlapping nodes. The size of the nodes is proportional to their weighted degree, while the size of links is proportional to their weights. The community structure is revealed using the SLPA detection algorithm.    
\label{f2}
}
\end{center}
\end{figure}

\noindent \textbf{Step 2:} form the sub-network based on the overlapping nodes and the hubs. The set of the overlapping nodes is denoted   $V_{o} = \{v_{1}^{o}, v_{2}^{o}, ...., v_{m}^{o}\} \subset V$ where $m$ is the number of the overlapping nodes. The set of hubs is denoted $V_{h} = \{v_{1}^{h}, v_{2}^{h}, ...., v_{t}^{h}\} \subset V$ where $t$ represents the numbers of hubs. The overlapping nodes and hubs network is formed from the union of the set of overlapping nodes and the set of hubs. It is obtained by removing all nodes that do not belong to one of the two sets. The overlapping nodes and hubs network is denoted $S(V_{y},E_{y})$, where $V_{y}= V_{o} \cup V_{h}$ and $E_{y}=\{(v_{i}^{y}, v_{j}^{y}) \setminus v_{i}, v_{j} \in V_{y}\}$ are respectively its set of nodes and edges. 
\autoref{f2} (b) illustrates the formation of this sub-network in the toy example shown in \autoref{f2} (a).\\

 Steps 3 and 4 are identical to steps 3 and 4 of the overlapping nodes ego backbone algorithm. \autoref{f2} (c) illustrates the overlapping nodes and hubs backbone when all the links with the lowest weights are discarded. In addition, \autoref{f2} (d) show the overlapping nodes and hubs backbone after setting the size threshold parameter $s$ to $0.4$.  

\subsubsection*{Computational Complexity}
The computational complexity is dominated by the community detection process. Indeed, if the community structure is unknown it must be uncovered. This can be done using a linear community detection algorithm \cite{jebablicomm}. For instance, it takes $O(N+L)$ for DEMON algorithm \cite{demon} while it takes $O(TL)$ for SLPA \cite{slpa} which is used in this work, where $N$ is the size of the network, $L$ is the number of its edges and $T$ is the maximum iteration of the algorithm. In the overlapping nodes ego backbone, the overlapping nodes and their immediate neighbors are extracted in $O(m)$, where $m$ is the number of the overlapping nodes.  In the overlapping nodes and hubs backbone, the nodes need to be sorted. This step can be done in $O(N)$.

\subsection*{Evaluation measures} 

Multiple evaluation measures are used in order to compare the extracted backbones. They are defined as follows:

\subsubsection*{Proportion of common nodes}   
This measure assesses how many elements belong to two different sets of the same size. It is defined as the fraction of the size of the intersection between the two sets divided by their size. The proportion of common nodes $A_{n}$ between two sets of nodes of the same size $n$ is defined as:
\begin{equation}
A_{n}= \frac{|X \cap Y|}{n} 
\end{equation}
Where $X$ and $Y$ are the sets of nodes of two different backbones. $n$ represents their size.\%.

\subsubsection*{Rank-biased overlap} 
The rank-biased overlap \cite{rbo} is an extension of the previous measure. It estimates how strongly two ranking sets are in harmony with each other. It computes the proportion of common elements between two rankings sets while gradually increasing their depths. The ranks are determined according to the weighted degree of nodes. Moreover, different weight is assigned to the overlap of each depth. Let's consider $X$ and $Y$ two rankings sorted in decreasing order. The rank-biased overlap $r$ of two ranking sets $X$ and $Y$ is given by:
\begin{equation}
r(X, Y) = \sum_{d} w_{d} * A_{d}
\end{equation}
Where $A_{d}$ designates the overlap between the two sets $X$ and $Y$ at depth $d$.
It is defined as follows:
\begin{equation}
A_{d} =  \frac{|X_{:d} \cap Y_{:d}|}{d}
\end{equation}
$X_{:d}$ denotes the set of elements (rankings) ranging from position $1$ to position $d$.  $|X_{:d} \cap Y_{:d}|$ represents the size of the overlap of both sets $X$ and $Y$ at depth $d$. 
Each weight at a given depth $d$ is represented by $w_{d}$ which is given by:
\begin{equation}
w_{d} = (1-p)*p^{d-1}  \text{\;\;\;\;where:\;\;} \sum_{d} w_{d} = 1
\end{equation}
The weights can be fixed via the parameter $p$. It ranges between $0$ and $1$. When it has a small value, more weight is given to the top elements of the sets. Thus, the tail of both sets gets a lower weight than the top. When $p$ approaches $1$, the comparison becomes deep in both sets.
Additionally, the values of the rank-biased overlap vary between $0$ and $1$. The sets $X$ and $Y$ are perfectly identical when the value is equal to $1$, while they are totally disjoint when the value is equal to $0$. 

\subsubsection*{Correlation} 
The correlation is used to evaluate the agreement between the set of nodes of two different backbones. We use two different types of correlation coefficients:

 
\textbf{Pearson correlation.} This correlation coefficient is a very familiar measure that assesses the association between two sets. Let's consider two sets $X$ and $Y$ of weighted degrees of nodes belonging respectively to the ordered set of nodes of two different backbones. The Pearson correlation $\rho$ between the two sets $X$ and $Y$ is defined as follows:
\begin{equation}
\rho(X,Y)=\frac{\sum_{i=1}^{n} (x_{i}-\bar{x})(y_{i}-\bar{y})}{\sqrt{\sum_{i=1}^{n} (x_{i}-\bar{x})^{2} \sum_{i=1}^{n} (y_{i}-\bar{y})^{2}}} 
\label{eq1}
\end{equation}
Where:
\begin{equation}
\overline{x}=\frac{\sum_{i=1}^{n} x_{i}}{n} \text{; \;\;\;\;\;} \overline{y}=\frac{\sum_{i=1}^{n} y_{i}}{n}
\end{equation}

\textbf{Kendall Tau correlation.} This correlation coefficient is usually used to measure the ranking consistency of two sets of nodes. Let's consider two ranked sets $X=(x_{1}, x_{2}, ..,x_{n})$ and $Y=(y_{1}, y_{2}, ..., y_{n})$ of size $n$. A pair of ranks $(x_{i}, y_{i})$ and $(x_{j}, y_{j})$ are considered concordant if $x_{i} > x_{j}$ and $y_{i} > y_{j}$ , or if $x_{i} < x_{j}$ and $y_{i} < y_{j}$. It is said discordant if $x_{i} > x_{j}$ and $y_{i} < y_{j}$, or if $x_{i} < x_{j}$ and $y_{i} > y_{j}$. The pair of ranks is neither concordant nor discordant if $x_{i} = x_{j}$ or $y_{i} = y_{j}$. Note that the ranks are computed according to the weighted degree. The Kendall Tau correlation $\rho_{\tau}$ between two ranking sets $X$ and $Y$ of size $n$ is given by:

\begin{equation}
\rho_{\tau}(X,Y)=\frac{n_{c}-n_{d}}{\frac{1}{2}n(n-1)}
\label{eq2}
\end{equation}

Where $n_{c}$ and $n_{d}$ stand for the number of concordant and discordant pairs, respectively. 
Both correlation coefficients belong to the interval [-1,1]. There is a positive monotonic association between two sets ($\rho > 0$), if the values of the two vectors tend to increase or decrease simultaneously. In addition, there is a negative monotonic association between two sets ($\rho < 0$), if the values of one vector tend to increase when the values of the other decrease. When $\rho$ is equal to $0$, there is an absence of a monotonic association between the two sets. Furthermore, the strength of the relation between the two sets is considered moderate if the coefficient ranges between 0.5 and 0.8, while it is significant if the coefficient ranges between 0.8 and 1.



\subsection*{Dataset description}
A set of real-world networks is collected from different fields (social, co-appearance, biological, technological and collaboration networks) to conduct a series of experiments. Their size ranges from hundreds to thousands of nodes and edges. 

\textbf{Zachary's karate club.} It is a well known social network describing the relationship between members of a karate club at a US university. Nodes are the members of the karate club while the links represent the friendship between two members. A number between $0$ and $8$ is associated with each pair of club members. It indicates the strength of the relationship between a pair of members.

\textbf{Intra-organisational.} This network is extracted from a consulting company. The nodes represent the $46$ employees. A link exists between two employees if they have exchanged information or advice requests. The link weights designate how often an employee has turned to his colleague for information or advice on work-related topics during a period of three months. They are differentiated on a scale from $0$ to $5$. 

\textbf{Freeman’s EIES.}  This network contains $48$ researchers working on social network analysis. The nodes represent researchers while the edges represent their personal relationships. The edges have a weight ranging between $0$ and $4$. The weight is equal to $0$ when a person is unknown to the researcher. It is equal to $1$ when a person the researcher has heard of but didn't meet. $3$ represents a friendship relationship while $4$ stands for a close personal friendship.

\textbf{Train bombing} This data set contains the network of suspected terrorists implicated in the train bombing which took place in Madrid on March 11, 2004. It was reconstructed from newspapers. The nodes represent terrorists while a link exists between two terrorists if there was contact between them. The weights indicate how strong their relationship was (a friendship or co-participating in training camps or previous attacks).

\textbf{Les Miserables.} This co-appearance network contains the characters in Les Miserables novel. The nodes are the characters and link exists between two characters if they co-appeared in the same chapter. The weights measure the number of such occurrences.

\textbf{Game of thrones} This co-appearance network contains the characters of the Game of Thrones series, more precisely in the book "A Storm of Swords". Nodes are the characters while a link exists between two characters if their names appeared within 15 words of each other in the text. The links are weighted by the number of these co-appearances.

\textbf{C.elegans Neural} The Caenorhabditis elegans worm's neural network contains 306 nodes representing neurons. An edge exists between two neurons if at least one synapse or gap junction exists between them. The weight is the number of synapses and gap junctions. 

\textbf{Facebook-like Social} It is an online social network of students at the University of California. The users included in this network are those who sent or received at least one message. The link weights of this network represent the number of characters in these messages.

\textbf{Facebook-like Forum} The Facebook-like Forum network does not focus on messages exchanged between users but their activity in the forum. This forum includes 899 users and 522 topics. The weights can be associated to the edges based on the number of messages that a user posted to a topic.

\textbf{US Power Grid} This undirected network represents the high voltage power grid of the Western States of the United States of America. Nodes are generators, transformers and substations while the edges are the high voltage lines connecting them. The weights indicate the total electrical load in different lines.

\textbf{Scientific Collaboration} This network contains authors of articles posted to the Condensed Matter section of arXiv E-Print Archive. Nodes are the authors and connections exist if they co-authored a paper. The weights are the number of joint papers.

\bibliography{mybibfile}

\section*{Author contributions}
All the authors contributed to designing the proposed method. ZG implemented the model and all the analyses. All authors participated in the formulation and writing of this paper. All authors approved the final manuscript.

\section*{Additional Information}
\textbf{Competing interests:} The authors declare that they have no competing interests.

\end{document}